\documentclass[twocolumn, pra, amssymb, superscriptaddress, aps, 
preprintnumbers,
amsmath,
floatfix%,
]{revtex4-1} 

\usepackage{multirow}
\usepackage{isotope}
\usepackage{graphicx}
\usepackage{xcolor}
\usepackage{braket}
\usepackage{enumitem}
\usepackage{pbox}

\begin{document}

\title{A neural network approach to running high-precision atomic computations}

\author{Pavlo Bilous}
\email{pavlo.bilous@mpl.mpg.de}
\affiliation{Max Planck Institute for the Science of Light, Staudtstr. 2, 91058 Erlangen, Germany}
\author{Charles Cheung}
\affiliation{Department of Physics and Astronomy, University of Delaware, Delaware 19716, USA}
\author{Marianna Safronova}
\affiliation{Department of Physics and Astronomy, University of Delaware, Delaware 19716, USA}

\date{\today}

\newcommand{\pb}[1]{\textcolor{red}{[Pavlo: #1]}}
\newcommand{\cc}[2][orange]{\textcolor{#1}{Charles: #2}}
\newcommand{\ccs}[2][orange]{\textcolor{#1}{#2}}
\newcommand{\MS}[2][red]{\textcolor{#1}{Marianna: #2}}

\begin{abstract}
Modern applications of atomic physics, including the determination of frequency standards, and the analysis of astrophysical spectra, require prediction of atomic properties with exquisite accuracy. For complex atomic systems, high-precision calculations are a major challenge due to the exponential scaling of the involved electronic configuration sets. This exacerbates the problem of required computational resources for these computations, and makes indispensable the development of approaches to select the most important configurations out of otherwise intractably huge sets. We have developed a neural network (NN) tool for running high-precision atomic configuration interaction (CI) computations with iterative selection of the most important configurations. Integrated with the established pCI atomic codes, our approach results in computations with significantly reduced computational requirements in comparison with those without NN support. We showcase a number of NN-supported computations for the energy levels of Fe$^{16+}$ and Ni$^{12+}$, and demonstrate that our approach can be reliably used and automated for solving specific computational problems for a wide variety of systems.
\end{abstract}

\maketitle

\section{Introduction\label{sec:intro}}

Accurate modeling of electronic correlations in atoms and ions is a ubiquitous problem in modern atomic physics. It is typically addressed using the configuration interaction (CI) approach \cite{Grant_book_2007}, which consists in expanding the wave function of the searched electronic states $\ket{\Psi} = \sum_\alpha c_\alpha \ket{\Phi_\alpha}$ in a fixed basis $\ket{\Phi_\alpha}$ (usually Slater determinants or configuration state functions). The unknown coefficients $c_\alpha$ for each state and the state energies $E_\alpha$ are then obtained as solutions to the eigenvalue problem for the Hamiltonian matrix $H_{\alpha\beta} = \braket{\Phi_\alpha | \hat H | \Phi_{\beta}}$. For accurate computations, the required basis set $\ket{\Phi_\alpha}$ can become huge, posing high demands both on the hardware and the atomic codes, which need to be highly efficient and parallelized. Even then, the required precision remains often beyond the computational feasibility.

In a recent work~\cite{Bilous_MLGRASP_PRL}, an algorithm using a neural network (NN) was demonstrated to tackle big CI computations for the GRASP2018 general relativistic atomic structure package \cite{GRASP}.
In this approach,  one large CI computation is replaced by a number of smaller ones on a subsequently growing sub-basis. The basis growth is managed by a NN classifier that receives the basis state quantum numbers as input.  In each iteration, it predicts the important basis states from the full set, i.e. those with weights $w_\alpha = |c_\alpha|^2$ exceeding a specified cutoff $x$.  They are then included in the CI computation that yields the energy and the coefficients $c_\alpha$. The latter are used to give the NN feedback on its prediction and retrain it. The ``importance'' cutoff $x$ is decreased from iteration to iteration, leading to convergence of the energy to its ``true'' value on the full set. In Ref.~\cite{Bilous_MLGRASP_PRL}, otherwise unfeasible computations for the ground state energy of the Re and Os atoms were performed. The results were compared with Ref.~\cite{Filianin_PRL_072502_2021}, where the CI problem was simplified down to a tractable scale using experimental data for electronic transition and excitation energies. The comparison showed a good agreement leading to the conclusion that the NN-supported algorithm yielded reliable results.

Unfortunately, in many cases, there are no experimental data to carry out such benchmarks as in Ref.~\cite{Bilous_MLGRASP_PRL}. This is especially the case in the domain of highly charged ions, which has yet to be thoroughly explored, but has recently been of great interest, due to its promising applications~\cite{Kozlov_RevModPhys_2018}. Therefore, it is necessary to have at hand an independent systematic \textit{ab initio} approach, which would ensure the validity of the NN-supported computations. Further analysis of the computations in Ref.~\cite{Bilous_MLGRASP_PRL} suggests that the required NN-related part was itself computationally demanding. The usual dense NN architectures failed, and only switching to a deep convolutional architecture tailored specifically to the structure of the input data (in Ref.~\cite{Bilous_MLGRASP_PRL}, configuration state functions) led to a stable behavior. Even on a GPU, the training of such NN took time comparable with the atomic computation in each iteration.

Here, we introduce significant algorithmic improvements and simplifications to the approach from Ref.~\cite{Bilous_MLGRASP_PRL}, making the computational overhead beyond the actual CI computations negligible. Instead of using the GRASP package \cite{GRASP} to implement CI, we use the 
pCI code package developed in Ref.~\cite{sym2021} to carry out extremely large-scale parallel CI computations. This package can be also used in combination with the coupled-cluster method and has 
demonstrated exceptional accuracy for a wide variety of many-electron systems, including neutral atoms, highly-charged ions, and negative ions \cite{2024Fe,2024Ni,2023Ti,2024Sn,2022Fe,2022Cr,2021La}. With the new implementation, the NN part needs only tens of minutes on a single CPU, which is negligible in comparison to the large-scale CI computation.  We also generalize the method from computations for only one electronic level (e.g. the ground state as in Ref.~\cite{Bilous_MLGRASP_PRL}) to many levels at once, i.e., yielding the electronic transition energies. The latter are undoubtedly of superior relevance in atomic physics and its applications. The computations performed in this work contained up to 17 levels at once. Importantly, we show how the obtained results can be verified using an alternative approach to perform large CI computations ``by parts'' without the NN. The developments in this work promote the NN-supported algorithm to a generic practical tool for high precision atomic computations.

As demonstration, we perform calculations of the energy levels on particularly large basis sets for the highly charged Fe$^{16+}$ and Ni$^{12+}$ ions. While Fe$^{16+}$ is directly relevant for understanding the astrophysical spectra, the Ni$^{12+}$ ion possesses a clock transition between its ground and second excited state, and thus has potential in metrology. These ambitious applications require high precision of a few cm$^{-1}$, which we achieve in our computations using the two complementary algorithms, with and without a NN. In these cases, obtaining a high level of accuracy requires the use of very large basis sets, with both large number of partial waves and large principal quantum numbers. The procedure to increase the basis set to numerical completeness at the required precision level is usually extremely costly in terms of computational resources. It also requires submission of a large number of multiple runs that limits such computations to a few exceptional cases. In the present work, we show that the NN-supported algorithm can solve the problem of the basis set increase with complete NN automation, as well as drastic reduction of the computer resource requirements, enabling future automated basis set upscaling for increased accuracy of computations.

The article is structured as follows. In Section~\ref{sec:ci}, we present the details of our CI computations and, in particular, the alternative approach to perform large CI computations by parts, without NN support. In Section~\ref{sec:nn_algo}, we describe in detail our NN-supported algorithm and highlight the improvements with respect to Ref.~\cite{Bilous_MLGRASP_PRL}. Section~\ref{sec:demos} contains the demonstration computations for the Ni$^{12+}$ and Fe$^{16+}$ ions. We finish the article by summarizing our conclusions in Section~\ref{sec:summary}.

\section{CI computations\label{sec:ci}}
All calculations are carried out using the CI method to correlate all valence electrons. We construct one-electron orbitals from solutions of the Hartree-Fock-Dirac (HFD) equations in the central field approximation. The basis set is designated by the highest principal quantum number for each included partial wave. For example, $17g$ means that all orbitals up to $n=17$ are included for the $spdfg$ partial waves. 

The CI many-electron wave function is obtained as a linear combination constructed from all distinct states and possessing a given total angular momentum and parity:
\begin{equation}
\ket{\Psi^k} = \sum_{\alpha} c_{\alpha}^k \ket{\Phi_\alpha}\;.
\end{equation}
Here the index $\alpha$ labels the involved Slater determinants $\ket{\Phi_\alpha}$, whereas $k$ enumerates the many-electron states. The energies and wave functions are determined from the time-independent many-electron Schr\"odinger equation 
\begin{equation}
\hat{H} \ket{\Psi^k} = E^k \ket{\Psi^k}\;.
\end{equation}

We construct one-electron basis orbitals from solutions of the HFD equations in the central field approximation. In general, we start by building core orbitals for the ground state configuration. Then we freeze those orbitals and construct valence orbitals for a few excited states. Virtual orbitals are then built from the HFD orbitals. A list of configurations defining the CI space is then obtained from all possible single (S) and double (D) excitations to any orbital in our basis set from a few selected reference configurations (typically the ground state and a few excited states). The Hamiltonian is then constructed and diagonalized to obtain the desired eigenvalues and eigenvectors. 

This procedure is completed for increasing basis sets (increasing principal quantum number $n$ or partial wave $l$) to identify the convergence pattern towards the most accurate results. This process is repeated until (1) either the energies have converged, or (2) the computation becomes too large for the available computational resources. Once the energies have converged as in (1), we consider our computations to be completed; otherwise, we treat (2) via a method optimizing and reducing the size of the CI space.

\subsection{Upscaling the basis set}
\label{noNN}
For computations of basis sets that cannot be run directly on currently available computational resources, we opt to reduce the size of the CI space by selecting only the most important configurations for subsequent calculations via some cutoff $x$. From a previously completed direct CI calculation, a configuration subspace corresponding to only the most important configurations can be generated. The importance of a configuration is determined by its weight $w_\alpha=|c_\alpha|^2$. An optimal cutoff $x$ is determined by balancing the number of configurations obtained from the cut, and the subsequent energy difference between the direct run and the cut run. This cutoff $x$ is typically chosen such that the resulting energy difference is minimized for any resulting energy level. This energy difference is then subtracted from the results of the subsequent CI calculation involving an increased principal quantum number $n$ or increased partial wave $l$.  We refer to this procedure of systematically increasing the basis set from a selected list of important configurations as \textit{upscaling the basis set}. 

As an example, we describe the process of upscaling the $17h$ basis set to the $20h$ basis set. In this case, we have completed the CI calculation for the $17h$ basis set, but $20h$ was not possible. Here, a cut $\log_{10}x=-10$ is used to reduce the size of the $17h$ basis set, with the resulting subspace consisting of only the most important configurations (we will denote it as $17h$\_cut). This cut reduces the size of the CI space from 242\,924 relativistic configurations (41\,071\,940 Slater determinants) to 89\,861 (13\,345\,491). We then construct a list of configurations obtained from increasing $17h$ to $20h$ and merge this list to the $17h$\_cut list. The resulting configuration list contains $17h\_\mathrm{cut}+(20h-17h)$. The CI procedure is then executed using this basis set to obtain the desired energies from the $20h$ basis set. However, it is important then to also subtract the energy difference due to the cut that was made to the $17h$ basis. So the final energy in the $20h$ calculation is $E_{20h}=E_{17h\_\mathrm{cut}+(20h-17h)}-E_{17h\_\mathrm{cut}-17h}$. In this way, we are able to approximate the exact CI results of the $20h$ basis by performing much smaller computations. We note that this procedure is still very time consuming and still requires both large memory and CPU allocations.

\section{NN-supported algorithm\label{sec:nn_algo}}

\begin{figure*}[ht!]
\centering
\includegraphics[width=\textwidth]{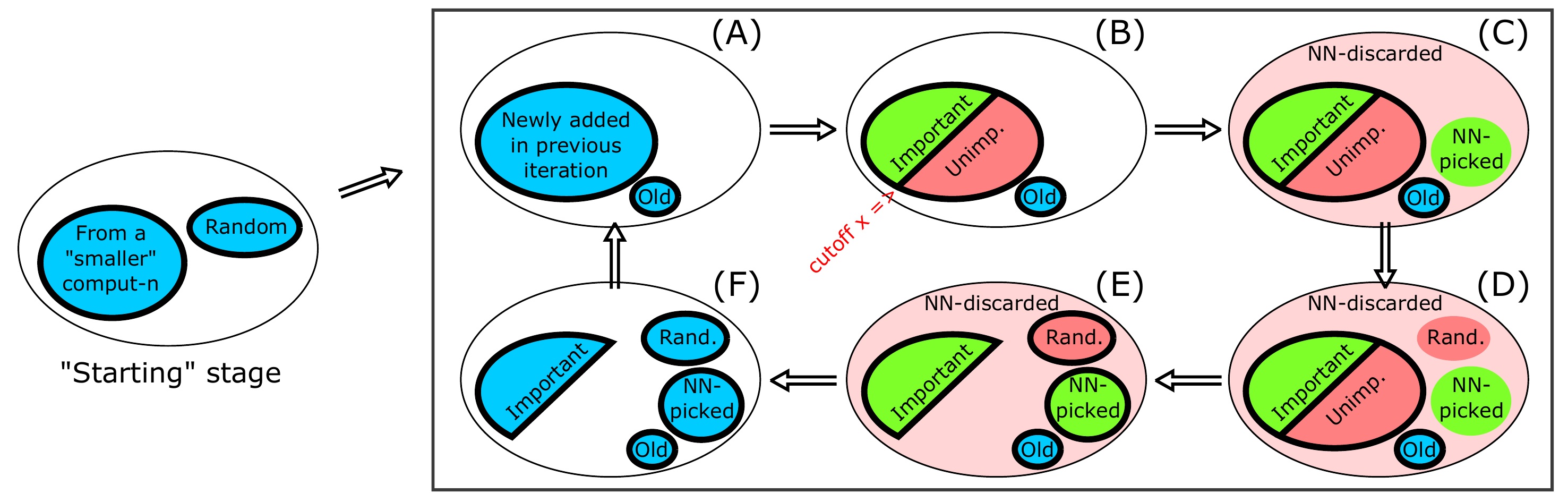}
 \caption{Schematic representation of the iterative NN-supported algorithm described in the text. The thin big ellipse denotes the full set of configurations (potentially intractable with a direct CI approach). The subsets shown with thick ellipses are all included in the current CI expansion. The blue color denotes configurations for which the CI expansion coefficients have been already evaluated. The green (red) color depicts the important (unimportant) configurations with respect to the current cutoff $x$ --- this is established either directly (if the CI coefficients are known) or using the NN. Note that the different shades of red are irrelevant. The algorithm begins with the ``starting'' stage (shown to the left) which assumes that a direct or NN-supported computation has been previously performed on a partial subset of the full set. A number of random configurations is additionally included, and a CI computation is performed yielding the coefficients for the included configurations. The algorithm then enters stage (A) of the NN-supported iterations (shown in the box to the right). The CI expansion contains some configurations either from the ``starting'' stage or the previous iteration [finished at stage (F)]. In the latter case, we distinguish between the configurations added directly in the previous iteration and the ones present in the CI expansion before --- the latter are denoted as ``Old''. At stage (B), we split the set of the ``newly added'' configurations into the important and unimportant classes with respect to the current cutoff $x$, and thus form a training set for the NN. The NN is trained and applied to classification of configurations from the pool at stage (C). At stage (D), a number of additional ``balancing'' configurations is randomly selected. At stage (E), the unimportant configurations are removed from the current CI expansion, whereas the NN-predicted and the ``balancing'' ones are included. At stage (F), the CI computation on the formed CI set is performed, finishing the iteration. The energy obtained at (F) is monitored to terminate the algorithm once convergence is achieved.}
\label{fig:algo}
\end{figure*}

In this section, we discuss a powerful alternative to the direct CI described above. We introduce our NN-supported iterative algorithm illustrated schematically in Fig.~\ref{fig:algo}, and highlight the improvements with respect to Ref.~\cite{Bilous_MLGRASP_PRL}. Here, we work not with the basis Slater determinants, but instead with relativistic configurations, which are constructed from the determinants. The weight of each configuration $\Gamma$ is obtained from the individual basis coefficients as $w_\Gamma^k = \sum_{\alpha \in \Gamma} |c_\alpha^k|^2$, where, as before, $k$ labels the considered energy level. We perform our computations for all levels of interest at once, and aggregate their weights in one weight as $w_\Gamma = \max_k w_\Gamma^k$. Given a large ``full'' set of configurations not tractable with the direct CI, the algorithm aims at forming an approximative CI subset of the most relevant configurations in the sense of their weight $w_\Gamma$. This approximative set (the ``current'' CI expansion) is improved with iterations by including further configurations from the remaining ``pool'', consisting of configurations from the full set which are currently not included in the CI expansion.

\subsection{``Starting'' stage\label{subsec:algo_start}}

At the beginning, we need to form an initial CI expansion, which serves as data for the initial NN training. We assume that before our algorithm starts, a computation on a smaller sub-basis of the full basis has already been performed. For instance, a computation for a smaller number of virtual orbitals can serve this purpose. Alternatively, the results of another NN-supported computation can be employed.

We construct the ``starting'' CI space by combining the initial CI space with a number of configurations chosen randomly from the remainder of the full basis. This set of randomly chosen configurations is needed for (a) the NN to explore the data not represented in the preceding computation, and for (b) mitigation of possible bias towards the ``important'' class of the training data. Note that the CI coefficients for the random selection are not yet known, and therefore a CI computation is performed at this step to calculate them. In case the obtained set is too big, a part of the sub-basis from the previous computation with weights below some cutoff $y$ can be omitted at this stage. We stress that this omission is done only for this particular CI computation step, since the aim is only evaluation of the CI coefficients for the random selection. The CI expansion obtained at the described "starting" stage is schematically shown to the left in Fig.~\ref{fig:algo}.

\subsection{Iterations}

We now enter the NN-supported iterations shown in Fig.~\ref{fig:algo} in the box to the right. At step (A), we assume that the CI expansion contains configurations from the ``starting'' stage described above or from the previous iteration [finished at (F)]. In the latter case, we distinguish between the configurations added directly in the previous iteration and the ones present in the CI expansion before (the latter are denoted as ``old''). Each iteration is characterized by its own importance cutoff $x$, which is decreased during the computation. At step (B), the newly added configurations in the CI expansion are distributed into the two importance classes with respect to the current cutoff $x$. At the next step (C), the NN is retrained on the obtained data and predicts new important configurations from the pool.

At step (D), we randomly select a number of configurations discarded by the NN, since the NN needs some feedback on them at the next iteration. This ``balancing'' is characterized by the ratio $r_\mathrm{bal}$ between the number of the randomly included configurations and the number of configurations classified by the NN as important. For larger $r_\mathrm{bal}$ values, the NN performs better in the next iteration, but the current CI expansion includes randomly selected and thus mostly irrelevant configurations, making the computational overhead larger. For small $r_\mathrm{bal}$ values, the computational overhead in the current iteration decreases, but in the next iteration, the NN may perform poorly, suggesting many irrelevant configurations. In Ref.~\cite{Bilous_MLGRASP_PRL}, the value $r_\mathrm{bal} = 1$ was used, whereas in the present work, the choice $r_\mathrm{bal} = 0.5$ is sufficient. We note that in the very last iteration, $r_\mathrm{bal} = 0$ can be chosen since no NN retraining follows.

At step (E), the current CI expansion is ``cleaned up'' by removing the configurations belonging to the ``unimportant'' class from step (D). We note that since the cutoff $x$ is iteratively decreased, and the NN state evolves with further training, these removed configurations may be classified by the NN as important and re-included in future iterations. The CI expansion is now enriched with the new configurations, and the CI computation is performed at step (F), yielding the energy levels and the CI expansion coefficients. The convergence of the energies is monitored and the algorithm is stopped once the targeted precision is achieved.

\subsection{NN architecture and training\label{sec:nn_archit_train}}

For each relativistic configuration, the input data consist of populations of the electronic orbitals. We convert this discrete input to a binary format by directly transforming the integer orbital populations to their binary code. Each binary digit is then treated as a feature in the transformed data. After this transformation, some of the resulting features may turn out to be trivial, i.e. constant (0 or 1) across the whole dataset, and are deleted. The NN output consists of 2 numbers interpreted as the probabilities of the input configuration to be important or unimportant. The probabilities are guaranteed to sum up to 1 by using the softmax activation function in the NN output layer.

With fixed input and output structure, there remains freedom in choice of the internal NN architecture. In contrast to Ref.~\cite{Bilous_MLGRASP_PRL}, where convolutional NNs were used, here, we employ the usual dense NNs~\cite{Goodfellow2016}. In all computations, we include 4 hidden layers, each with a few tens to a few hundreds of neurons, resulting in a number of trainable parameters of the order of $10^5$ (concrete sizes of the input and the NN layers will be specified for each example separately). The NN is trained to minimize the categorical cross-entropy loss using the Adam optimization algorithm~\cite{kingma2017adam}. The training is stopped based on the NN performance monitored on 20\% of the data held out randomly from the training set. The described functionality is leveraged using the Python library TensorFlow~\cite{TensorFlow2015}.

\subsection{Algorithm improvements\label{sec:improv}}

Here, we explicitly highlight the main improvements with respect to the algorithm developed in Ref.~\cite{Bilous_MLGRASP_PRL}.

\subsubsection{Grouping basis states in relativistic configurations}

A crucial change with respect to Ref.~\cite{Bilous_MLGRASP_PRL} consists in treating the basis states in groups of Slater determinants corresponding to relativistic configurations instead of considering them individually. The weight of a configuration $\Gamma$ is the sum of the individual weights of the participating basis states $w_\Gamma^k = \sum_{\alpha \in \Gamma} |c_\alpha^k|^2$. In this case, all information distinguishing the basis states within a single relativistic configuration is discarded. This has two major advantages. This significantly simplifies the input structure and reduces the number of features (the dataset ``width'') without harming the quality of the results, and this strongly reduces the amount of data (the dataset ``length'') and is advantageous for large computations. The resulting data are well compatible with the commonplace dense NN architecture. This leads to a stable and well reproducible computation flow and results, without the need of introducing a deep convolutional block as in Ref.~\cite{Bilous_MLGRASP_PRL}.

Another advantage of switching to relativistic configurations is a unification of computations performed in the basis of Slater determinants (used in the present work) and configuration state functions (used in Ref. \cite{Bilous_MLGRASP_PRL}). Specifically for our atomic code package~\cite{sym2021}, this choice also corresponds to the standard input/output mechanism.

\subsubsection{Multilevel optimization}

In this work, we obtain all energy levels of interest in one computation, rather than optimizing them one by one, as described in Ref.~\cite{Bilous_MLGRASP_PRL}. We achieve this by switching from individual weights $w_\Gamma^k$ for each $k$-th energy level to an ``aggregated'' weight $w_\Gamma = \max_k w_\Gamma^k$ for each relativistic configuration $\Gamma$. With this choice, a configuration becomes important if it is important for at least one energy level. We note that other aggregation approaches tailored to a specific computation can be employed here.

A special challenging situation occurs in computations for the Ni$^{12+}$ ion due to an extreme closeness of its 1st and 2nd excited states, lying approximately 600 cm$^{-1}$ apart. Optimization for each level separately in an independent computation based on the algorithm from Ref.~\cite{Bilous_MLGRASP_PRL} led to a wrong ordering of the levels and difficulties in their identification. In our improved algorithm, all levels are optimized, preventing such mixing due to poor quality energies. This makes our approach practical also in such peculiar situations.

\subsubsection{Starting from a prior computation\\(direct or NN-supported)}

In our algorithm, each iterative computation starts not from scratch as in Ref. \cite{Bilous_MLGRASP_PRL}, but from the output of a prior (direct or NN-supported) computation. This approach is compatible with the standard basis expansion procedures in atomic computations,  e.g. increasing the highest principal quantum number $n$ for the included virtual orbitals. More importantly, the information from the previous computation is reused for the starting training of the NN classifier. This resolves the issue of the ``starting'' iteration, in which the NN is not yet trained, and thus unable to predict important basis states. In Ref.~\cite{Bilous_MLGRASP_PRL}, random basis states were drawn from the full set and included in the computation as the initial training dataset. A great disadvantage of such approach is strongly unbalanced data due to the lack of important basis states in the random selection. This severely contributes to instabilities, which had to be tackled in Ref.~\cite{Bilous_MLGRASP_PRL} at a high computational cost, but are circumvented in this work. With this improvement, there is also no need to manually select a so-called ``primary set'' to be explicitly included in each iteration as in Ref.~\cite{Bilous_MLGRASP_PRL}.

\section{Demonstration computations\label{sec:demos}}

In this section, we demonstrate our approach by performing computations on particular basis sets for:
\begin{enumerate}[label=\Alph*)]
\item 5 lowest Ni$^{12+}$ levels (belonging to the ground state fine structure);
\item 5 lowest even states of the Fe$^{16+}$ ion;
\item 17 lowest odd states of the Fe$^{16+}$ ion.
\end{enumerate}
These examples cover the most relevant aspects of the NN-supported computations. We concentrate here on obtaining high precision results using the improved version of the algorithm, and refer to Ref.~\cite{Bilous_MLGRASP_PRL} for further demonstration and discussions of the basic procedure.

\subsection{Numerical results \label{sec:demo_num_res}}

\subsubsection{Five lowest states of Ni$^{12+}$}

We start with a demonstration of our NN-supported approach for computations of the five lowest energy levels of Ni$^{12+}$, which all belong to the ground state fine structure. Such ions with optical narrow transitions are of particular importance to the development of highly-charged-ion clocks \cite{Kozlov_RevModPhys_2018}. 
We consider Ni$^{12+}$ with a closed core [$1s^2\,2s^2\,2p^6$] and ground state configuration $3s^2\,3p^4$ with 6 valence electrons. The core can be accounted for using the coupled-cluster approach as in e.g. Ref~\cite{2020Cf}, so here we focus on making the 6-electron CI computation complete with respect to the basis set convergence.

The full basis is constructed by SD excitations from the reference electronic configurations $3s^2\,3p^4$,  $3s\,3p^4\,3d$ and $3s^2\,3p^2\,3d^2$ to the orbitals up to $22spdfgh20ikl$, resulting in 862\,788 relativistic configurations (208\,827\,180 Slater determinants). For this set of orbitals, the number of features after the binary transformation of the input data (see Section~\ref{sec:nn_archit_train}) is 575. We use 4 hidden NN layers of sizes 150, 75, 45, 20, resulting in 102\,107 trainable NN parameters. We perform a full CI computation on a restricted subset of 171\,644 configurations (24\,064\,676 Slater determinants) obtained by limiting the basis set to $17spdfg$, and use it as the starting point for the NN-supported computation.

In Table~\ref{tab:ni}, we compare the energies of the lowest 5 levels evaluated using the direct CI computation (performed using the basis upscaling technique described in Section~\ref{noNN}) and obtained for iterations of the NN-supported approach. The latter are labeled by their cutoff value $x$; the ``starting'' stage is also presented. The final energy value in the NN-supported algorithm is obtained by exponential extrapolation of the values in the iterations (discussed in more detail later in the text). The energies $E^k_\mathrm{direct}$ from the direct CI computation are shown in Hartree units, whereas the energies obtained in the NN-supported algorithm are counted from the corresponding $E^k_\mathrm{direct}$ values and are measured in cm$^{-1}$. It is seen that in the subsequent iterations with decreasing importance cutoff $x$, the energies $E^k_i$ are refined and converge eventually to the values $E^k_\mathrm{direct}$ with the targeted precision of a few cm$^{-1}$. The resulting basis set from the NN consists of 145\,490 configurations, a factor of approximately 6 times less than that of the full basis set. Note that the energy differences between the levels obtained using the two computation methods agree within an error of 1 cm$^{-1}$ smaller than for the individual levels. We attribute this cancellation effect to the similarity of the considered levels, since they all belong to the fine structure of the same electronic state.

\begin{table*}[ht!]
\begin{tabular}{|c|c||c||c|c|c|c|c|c|c|}
\hline
\multirow{ 2}{*}{Level $k$} &
\multirow{ 2}{*}{Configuration} &
\multirow{ 2}{*}{$E^k_\mathrm{direct}$, Hartree} &
\multicolumn{7}{|c|}{$E^k_i - E^k_\mathrm{direct}$, cm$^{-1}$} \\\cline{4-10}
& & & Start & -10.0 & -10.5 & -11.0 & -11.5 & -12.0 & \textbf{Final} \\\hline
\hline
0 & $3s^2 3p^4$\,$^3P_2$ & -1457.78467907 & 1294.4 & 257.3 & 88.8 & 35.5 & 18.4 & 9.3 & \textbf{3.8}   \\\hline
1 & $3s^2 3p^4$\,$^3P_1$ & -1457.69624247 & 1266.1 & 195.4 & 73.5 & 30.6 & 16.4 & 8.2 & \textbf{2.7}   \\\hline
2 & $3s^2 3p^4$\,$^3P_0$ & -1457.69337789 & 1309.3 & 256.3 & 91.6 & 37.0 & 20.1 & 9.9 & \textbf{3.6}   \\\hline
3 & $3s^2 3p^4$\,$^1D_2$ & -1457.56972999 & 1458.1 & 238.1 & 88.1 & 35.8 & 19.2 & 9.7 & \textbf{3.6}   \\\hline
4 & $3s^2 3p^4$\,$^1S_0$ & -1457.33747811 & 1506.4 & 248.6 & 94.4 & 39.5 & 21.8 & 11.1 & \textbf{3.8}   \\\hline
\end{tabular}
\caption{Comparison of the 5 lowest energy levels of Ni$^{12+}$ between completed direct CI computation (performed on basis parts using the basis upscaling technique) and iterations of the NN-supported algorithm. The final approximative CI expansion has a size of 145\,490 relativistic configurations.\label{tab:ni}}
\end{table*}

\subsubsection{Five even energy levels of Fe$^{16+}$}

Next, we compute the 5 lowest even-parity energy levels of the highly charged Fe$^{16+}$ ion. Here, we consider all 10 electrons to be active, and allow SD excitations to vacancies in all orbitals up to $24h$ starting from 3 reference configurations $1s^2\,2s^2\,2p^6$, $1s^2\,2s^2\,2p^5\,3p$, and $1s^2\,2s\,2p^6\,3s$. This leads to 649\,195 relativistic configurations, corresponding to 48\,174\,193 Slater determinants, which is beyond our computational capabilities for a direct CI computation on the full set. The number of data features is in this case 469, and 4 hidden NN layers of size 469, 469, 234 and 117, respectively, were used, resulting in 578\,571 trainable NN parameters. The ``starting'' set is constructed as follows. We begin with a subset of 204\,487 relativistic configurations obtained from the full basis by restricting the virtual orbitals by $17g$. We perform a full CI computation on this set, and then omit all non-relativistic configurations with weights below $10^{-11}$. The resulting subset consists of 133\,276 relativistic configurations, corresponding to 9\,148\,479 Slater determinants. In Table~\ref{tab:fe_even}, we summarize the results of our computations in the same manner as in the previous section. The final approximative CI expansion contains 207\,604 relativistic configurations.

\begin{table*}[ht!]
\begin{tabular}{|c|c||c||c|c|c|c|c|c|c|c|}
\hline
\multirow{ 2}{*}{Level $k$} &
\multirow{ 2}{*}{Configuration} &
\multirow{ 2}{*}{$E^k_\mathrm{direct}$, Hartree} &
\multicolumn{8}{|c|}{$E^k_i - E^k_\mathrm{direct}$, cm$^{-1}$} \\\cline{4-11}
& & & Start & -9.5 & -10.0 & -10.5 & -11.0 & -11.5 & -12.0 & \textbf{Final} \\\hline
\hline
0 & $2p^6$\,$^1S_0$ & -1148.40541049 & 2074.8 & 523.6 & 294.5 & 133.1 & 31.0 & 5.0 & 1.5 & \textbf{0.2}   \\\hline
1 & $2p^5 3p$\,$^3S_1$ & -1120.64481897 & 1829.3 & 672.7 & 453.0 & 196.5 & 48.8 & 11.4 & 5.0 & \textbf{2.4}   \\\hline
2 & $2p^5 3p$\,$^3D_2$ & -1120.51593721 & 1800.5 & 523.6 & 330.7 & 141.5 & 33.0 & 6.1 & 2.3 & \textbf{0.9}   \\\hline
3 & $2p^5 3p$\,$^3D_3$ &  -1120.45669580 & 1798.6 & 522.9 & 298.6 & 129.1 & 28.5 & 5.3 & 1.9 & \textbf{0.6}   \\\hline
4 & $2p^5 3p$\,$^1P_1$ & -1120.41525162 & 1827.4 & 581.6 & 361.9 & 157.6 & 36.9 & 7.1 & 3.0 & \textbf{1.5}   \\\hline
\end{tabular}
\caption{Comparison of the 5 lowest even energy levels of Fe$^{16+}$ between completed direct CI computation (performed on basis parts using the basis upscaling technique) and iterations of the NN-supported algorithm. The final approximative CI expansion has a size of 207\,604 relativistic configurations.\label{tab:fe_even}}
\end{table*}

\subsubsection{17 odd energy levels of Fe$^{16+}$}

We proceed now to the lowest 17 odd states of the Fe$^{16+}$ ion, and target a challenging number of energy levels computed at the same time. As in the preceding section, we construct the full basis by allowing SD excitations for all 10 electrons to higher orbitals up to $24h$. The reference configurations are now $1s^2\,2s^2\,2p^5\,3s$, $1s^2\,2s^2\,2p^5\,3d$, and $1s^2\,2s\,2p^6\,3p$. This gives 947\,766 relativistic configurations (94\,527\,257 Slater determinants) in the full basis set. The same number of features is present in the NN input data and the same NN structure was used here as in the computations for the even Fe$^{16+}$ levels described in the previous section. We perform a prior full CI computation on a sub-basis of 296\,993 relativistic configurations (24\,115\,133 Slater determinants) constructed in the same way as for the even Fe$^{16+}$ levels. The results are shown in Table~\ref{tab:fe_odd}.  The final approximative CI expansion consists of 351\,452 relativistic configurations.
\begin{table*}[ht!]
\begin{tabular}{|c|c||c||c|c|c|c|c|c|c|c|}
\hline
\multirow{ 2}{*}{Level $k$} &
\multirow{ 2}{*}{Configuration} &
\multirow{ 2}{*}{$E^k_\mathrm{direct}$, Hartree} &
\multicolumn{8}{|c|}{$E^k_i - E^k_\mathrm{direct}$, cm$^{-1}$} \\\cline{4-11}
& & & Start & -9.5 & -10.0 & -10.5 & -11.0 & -11.5 & -12.0 & \textbf{Final} \\\hline
\hline
0 & $2s^2 2p^5 3s$\,$2$ & -1121.76048561 & 2263.9 & 866.8 & 658.0 & 252.9 & 71.6 & 17.3 & 7.2 & \textbf{3.0}   \\\hline
1 & $2s^2 2p^5 3s$\,$^3P_1$ & -1121.69077098 & 2282.1 & 877.0 & 667.8 & 255.5 & 72.6 & 17.7 & 7.3 & \textbf{3.0}   \\\hline
2 & $2s^2 2p^5 3s$\,$^1P_1$ & -1121.25287031 & 2175.6 & 877.2 & 661.8 & 264.7 & 74.2 & 18.7 & 8.0 & \textbf{3.4}   \\\hline
3 & $2s^2 2p^5 3d$\,$^3P_1$ & -1118.92128256 & 2286.3 & 829.2 & 594.5 & 245.5 & 70.2 & 14.4 & 5.0 & \textbf{0.9}   \\\hline
4 & $2s^2 2p^5 3d$\,$^3P_2$ & -1118.85451593 & 2299.9 & 837.6 & 591.5 & 253.3 & 67.5 & 15.0 & 4.8 & \textbf{0.2}   \\\hline
5 & $2s^2 2p^5 3d$\,$^3F_4$ & -1118.85179744 & 2329.5 & 853.0 & 599.7 & 260.1 & 68.1 & 16.2 & 5.1 & \textbf{-0.0}   \\\hline
6 & $2s^2 2p^5 3d$\,$^3F_3$ & -1118.82459374 & 2362.7 & 859.1 & 621.7 & 254.6 & 74.2 & 15.6 & 5.7 & \textbf{1.5}   \\\hline
7 & $2s^2 2p^5 3d$\,$^1D_2$ & -1118.76136360 & 2363.2 & 858.6 & 618.1 & 255.5 & 72.7 & 15.5 & 5.5 & \textbf{1.1}   \\\hline
8 & $2s^2 2p^5 3d$\,$^3D_3$ & -1118.72173088 & 2375.2 & 857.8 & 605.0 & 260.3 & 67.9 & 16.0 & 4.9 & \textbf{-0.1}   \\\hline
9 & $2s^2 2p^5 3d$\,$^3D_1$ & -1118.55136331 & 2371.3 & 855.5 & 608.8 & 259.8 & 71.0 & 15.9 & 5.1 & \textbf{0.2}   \\\hline
10 & $2s^2 2p^5 3d$\,$^3F_2$ & -1118.36202727 & 2308.3 & 838.7 & 598.3 & 267.1 & 77.5 & 17.0 & 5.5 & \textbf{0.0}   \\\hline
11 & $2s^2 2p^5 3d$\,$^3D_2$ & -1118.33212227 & 2296.9 & 855.9 & 587.0 & 264.7 & 72.7 & 16.3 & 5.6 & \textbf{0.6}   \\\hline
12 & $2s^2 2p^5 3d$\,$^1F_3$ & -1118.31283417 & 2317.1 & 866.8 & 596.1 & 271.4 & 76.5 & 17.5 & 6.2 & \textbf{0.7}   \\\hline
13 & $2s^2 2p^5 3d$\,$^1P_1$ & -1118.05785037 & 2373.0 & 859.3 & 607.3 & 267.8 & 76.2 & 17.2 & 5.5 & \textbf{-0.1}   \\\hline
14 & $2s 2p^6 3p$\,$^3P_1$ & -1115.59428826 & 2317.9 & 971.2 & 731.6 & 306.2 & 89.5 & 24.7 & 8.3 & \textbf{0.0}   \\\hline
15 & $2s 2p^6 3p$\,$^3P_2$ & -1115.50892176 & 2275.1 & 939.9 & 712.2 & 286.2 & 88.2 & 23.1 & 8.1 & \textbf{0.8}   \\\hline
16 & $2s 2p^6 3p$\,$^1P_1$ & -1115.43721159 & 2309.4 & 964.9 & 730.6 & 293.6 & 90.6 & 24.3 & 8.8 & \textbf{1.2}   \\\hline
\end{tabular}
\caption{Comparison of the 17 lowest odd energy levels of Fe$^{16+}$ between completed direct CI computation (performed on basis parts using the basis upscaling technique) and iterations of the NN-supported algorithm. The final approximative CI expansion has a size of 351\,452 relativistic configurations.\label{tab:fe_odd}}
\end{table*}

\subsection{Extrapolation of energies}

\begin{figure*}[ht!]
\centering
\includegraphics[width=\textwidth]{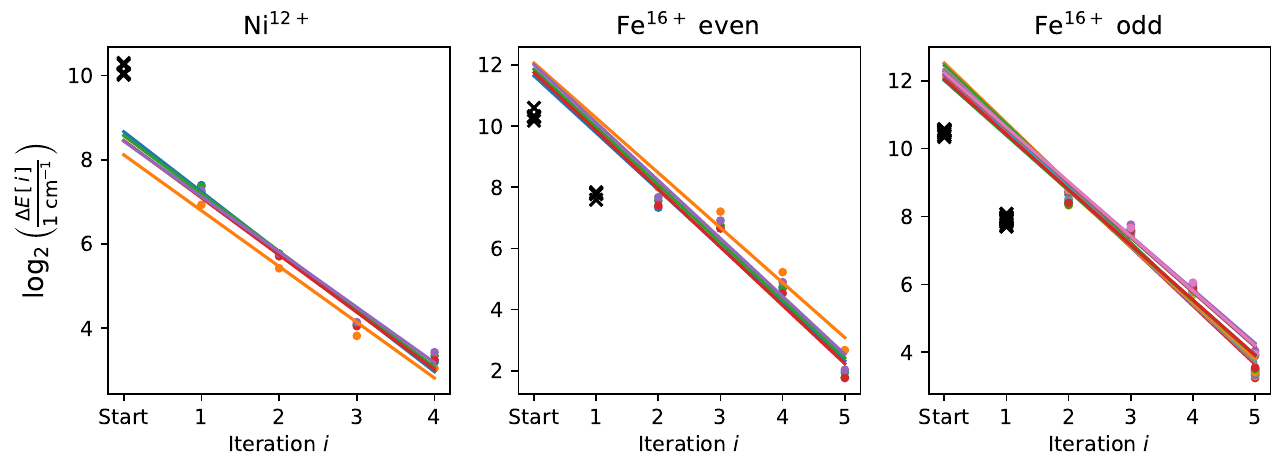}
 \caption{Extrapolation procedure for energies obtained in the iterations of the NN-supported algorithm for the three computations described in Section~\ref{sec:demo_num_res} is shown in the corresponding panels. Here, $\Delta E[i] = E_i - E_{i + 1}$ is the change of the level energies obtained in adjacent iterations indexed as $i$ and $i + 1$. The first iterations in which the pattern given by Eq.~(\ref{eq:extrap_assumption}) is not yet established, are marked with a black cross and omitted in extrapolation. As shown in color, each energy level $k$ has its own value $\Delta E^k[i]$, which all behave very similarly in the context of the extrapolation procedure.}
\label{fig:extrap}
\end{figure*}

In the cases described above, and shown in Tables~\ref{tab:ni}---\ref{tab:fe_odd}, the final level energies are obtained via exponential extrapolation of the values yielded in each NN-supported iteration. This procedure is based on the observation that after a few first iterations, the energy change $\Delta E[i] = E_i - E_{i + 1}$ between adjacent iterations starts decreasing by a factor
\begin{equation}\label{eq:extrap_assumption}
\frac{\Delta E[i]}{\Delta E[i + 1]} = \kappa \;,
\end{equation}
which is approximately constant. Once $\kappa$ is obtained numerically from linear fitting of $\log
\left(\Delta E[i]\right)$ as shown in Fig.~\ref{fig:extrap}, this allows us to formally sum up all energy changes $\Delta E[i]$ beyond our iterations and obtain the final result. From Fig.~\ref{fig:extrap}, it is seen that in the few first iterations, the described pattern is not yet established. These apparent outliers are not included in the extrapolation procedure. Though the assumption given by Eq.~(\ref{eq:extrap_assumption}) is satisfied not at all precisely, our demonstrations show that it is sufficient to improve the energy precision and avoid further costly iterations on a larger CI expansion. This is especially the case for the Ni$^{12+}$ example, since all energy levels belong to the fine structure of the same electronic (ground) state and behave similarly in the NN-supported computation. We stress that the extrapolation approach is applicable only to the energies and does not influence the CI expansion, e.g. the wave function, itself.

\subsection{NN training}

\begin{figure*}[ht!]
\centering
\includegraphics[width=\textwidth]{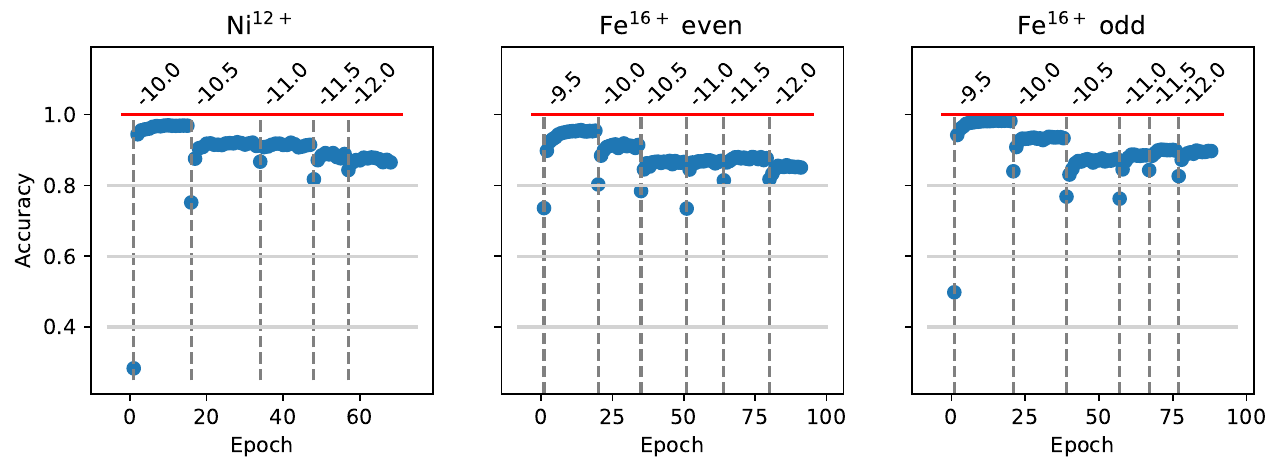}
 \caption{Evolution of the NN classification accuracy with the training epochs for the three computations described in Section~\ref{sec:demo_num_res} is shown in the corresponding panels. For each case, iterations of the NN-supported algorithm are shown separated by the vertical dashed lines and labelled by their importance cutoff $x$ at the top.}
\label{fig:nn_training}
\end{figure*}

Here we discuss the evolution of the NN performance with the training epochs. The three panels in Fig.~\ref{fig:nn_training} represent the NN training process for the three computations described in the previous section. For each case, we show the iterations of the NN-supported algorithm separated by vertical dashed lines and labelled by the importance cutoff $\log_{10} x$ at the top of each panel. The NN performance is measured by the classification accuracy, i.e. the fraction of the configurations classified correctly. It is evaluated prior to the training on the whole training data, and after each epoch on 20\% configurations held out from the training set. This accuracy is monitored to early-stop the NN training for the current iteration of the NN-supported algorithm if no progress is achieved. In this work, we use ``early stopping with patience'' and terminate the training not immediately after the epoch with the lower accuracy, but in case 5 further epochs have not led to an improvement. The NN is then reset back to its state immediately after the epoch in which the best accuracy was achieved.

The plots in Fig.~\ref{fig:nn_training} show the typical patterns of the NN training in our algorithm. In each iteration, the starting accuracy prior to training (shown by the points lying directly on the vertical dashed lines) is lower since the NN is either yet untrained (in the very first iteration) or has switched from an iteration with a different importance cutoff $x$. The accuracy then grows strongly already after the first training epoch and is refined in further epochs until the best NN performance is achieved. The slight accuracy decline in the last 5 epochs is attributed to the early stopping method used with patience=5. Although these epochs are shown in Fig.~\ref{fig:nn_training}, they do not participate in the NN evolution, since it is reset to its best state achieved in the iteration. We note that in the first iteration, higher accuracy is typically achieved due to training on the data obtained in an independent (direct or NN-supported) CI computation. This contrasts with the training procedure in Ref.~\cite{Bilous_MLGRASP_PRL}, where the first NN training was performed on a random selection from the pool, leading to unstable NN performance and necessity to switch to the convolutional architecture.

\subsection{Computational resources}

We now shift attention to the amount of used computational resources, as well as the total execution time of the calculations. Since both our NN-supported algorithm and the basis upscaling approach without NN involve many internal iterations, we stick to the representative case of Ni$^{12+}$ for comparison of the required computational effort. In Table~\ref{tab:runtimes}, we list the number of processors and required amount of memory per processor, as well as the runtime for each calculation for the respective basis sets. All computations are done with the maximum amount of available computational resources at the time. We find a total speedup of about 4 times of the NN-supported approach with respect to the direct CI with basis upscaling. Note that this only takes into account the runtime of the actual calculations with the CI atomic code, excluding any time to prepare basis sets and construct configuration lists (both with and without NN). The total time it takes to prepare the calculations is much lower in the case of the NN-supported approach, since here much fewer CI runs are performed in total, whereas the NN operation introduces no significant computational overhead.

\begin{table}[ht!]
\begin{tabular}{|c|c|c|c|}
\hline
\multicolumn{4}{|c|}{Direct CI} \\ \hline
Basis & num\_procs & mem\_per\_core & time (hr) \\
\hline
$17g$   & 1124 &  8.3 &    4  \\ \hline
$22g$   & 1124 & 17.7 &   13  \\ \hline
$17h$   & 1124 & 15.1 &  9.5  \\ \hline
$20h$   &  704 & 14.4 & 10.5  \\ \hline
$22h$   &  850 & 13.1 &   10  \\ \hline
$17i$   &  840 & 29.4 &   27  \\ \hline
$20i$   &  704 & 25.2 &   25  \\ \hline
$17k$   &  840 & 17.8 &   19  \\ \hline
$20k$   &  810 & 28.3 &   34  \\ \hline
$17l$   &  640 & 29.7 &   47  \\ \hline
$18l$   & 1350 & 15.4 &  10.5 \\ \hline
$19l$   & 1056 & 20.4 &  15.5 \\ \hline
$20l$   & 1000 & 25.3 &   24  \\ \hline 

$17h^a$ &  800 &  3.7 &  2.5  \\ \hline
$20h^a$ &  640 &  5.9 &    2  \\ \hline
$17i^a$ &  900 &  5.9 &    4  \\ \hline
$20i^a$ &  950 &  9.3 &  4.5  \\ \hline
$20i^b$ &  900 &    7 &    2  \\ \hline
$17k^a$ &  950 &  8.8 &  6.5  \\ \hline
$17l^b$ & 1350 &    8 &    2  \\ \hline
$18l^b$ &  576 & 11.5 &   5.5 \\ \hline
$19l^b$ &  660 & 13.9 &   5.5 \\ \hline
Total   &  ---    & ---  & 283.5 \\ \hline
\hline
\multicolumn{4}{|c|}{NN-supported CI} \\ \hline
Iteration & num\_procs & mem\_per\_core & time (hr) \\
\hline
Start   &  640 & 21.7 &   2.5 \\ \hline
-10     &  640 & 21.1 &   2.2 \\ \hline
-10.5   &  640 & 22.4 &   5.3 \\ \hline
-11     &  640 & 24.7 &   11.1 \\ \hline
-11.5   &  640 & 27.8 &  19.7 \\ \hline
-12     &  640 & 31.5 &  30.5 \\ \hline
Total   &  ---  & --- &  71.2    \\ \hline

\end{tabular}
 \caption{Computational resources and execution time for each computation required to obtain the final energies for Ni$^{12+}$. For each computation, the ``num\_procs'' column lists the number of CPU cores, and the ``mem\_per\_core'' column lists the total amount of required memory in GiB. The superscripts ``$a$'' and ``$b$'' indicate runs done with a cutoff $\log_{10}x=-10$ and $\log_{10}x=-9$, respectively.}
 \label{tab:runtimes}
\end{table}

\section{Conclusions\label{sec:summary}}
We have built upon the NN-based approach developed in Ref~\cite{Bilous_MLGRASP_PRL} and introduced significant developments and simplifications for its use with the pCI code package \cite{sym2021}. The improved algorithm operates on groups of Slater determinants corresponding to relativistic electronic configurations. The computation is performed for many levels at once, and starts from another (direct or NN-supported) computation. These improvements allow us to refrain from a computationally demanding convolutional architecture, as in Ref.~\cite{Bilous_MLGRASP_PRL}, and use a simple dense NN. We have successfully demonstrated the validity and accuracy of the results obtained using our NN-supported algorithm compared to the direct large-scale CI calculations for energy levels of the highly charged Ni$^{12+}$ and Fe$^{16+}$ ions. For the cases considered in this work, a direct CI computation on the full basis is not feasible. Instead, all direct CI calculations (i.e. without NN) were performed on partial sets using the basis upscaling technique. It has been shown that the NN-supported computation can be completed with significantly reduced execution time in comparison with basis upscaling. The demonstrated NN-supported basis set convergence procedure can now be automated to provide accurate results for a wide variety of systems for astrophysics and the development of atomic clocks.

\section*{Acknowledgements\label{sec:summary}}
We thank Florian Marquardt, Ian Grant, Per J\"onsson, Adriana P\'alffy and Chunhai Liu for helpful discussions. This work was supported by the US NSF Grant No. PHY-2309254 and US Office of Naval Research Grant No. N00014-20-1-2513.
The calculations in this work were done through the use of Information Technologies resources at the University of Delaware, specifically the high-performance Caviness and DARWIN computer clusters.  

\section*{References}
\bibliography{refs}

\end{document}